\magnification1200

{\bf MANY TIME INTERPRETATION OF THE QUANTUM
MEASUREMENT PROCESS}

\medskip

\centerline{Miroljub Dugi\' c}

\medskip

\centerline{\it Department of Physics, Faculty of Science}

\smallskip

\centerline{\it P.O.Box 60, 34 000
Kragujevac, FR Yugoslavia}

\centerline{e--mail : dugic@uis0.uis.kg.ac.yu}

\bigskip

{\bf Abstract} : This is a short version  of the, so-called, Many Time 
Interpretation (MTI) of the quantum measurement process. MTI does not
involve any new hypothesis, but rather represents a suggestion for
"re-reading" of the positive statements
of the existing measurement theories.
Quite shortly : Instead of the "indeterminism" with regard to the
unique Time, in nonrelativistic quantum measurement theory MTI 
establishes "determinism",
but with regard to the sets of the different (local) Times.

The very basis of MTI is a semi-epistemological analysis of 
the concept of physical time (Time) in classical, nonrelativistic 
physics. Particularly,
we offer an operational definition of Time,
which "establishes" Time as an "ordering principle" of the 
classical-physics world.

When applied
to the analysis of the quantum measurement process, 
the above mentioned definition 
strongly suggests the main interpretational rule of MTI: 
each particular (single) "object" in due course of quantum
measurement represents an object of stochastic change
(choice) of Time. Thus each particular "object" expiriences its own,
local Time, which is as objective (real) for the actual 
"object", as the "macroscopic" Time is objective in classical 
physics. 

	However, the analysis concerning the "object" alone appears
somewhat naive. The full analysis refers to the
composite systems, "object + apparatus (O+A)", and "object + 
apparatus + environment (O+A + E)". This way one obtains justification 
of the analysis concerning the "object" alone, however implying 
some nontrivial physical notions. Particularly, the given axiomatization
of MTI leads to : (i) Recognizing the amplification process as the
fundamental "part" of the measurement process, (ii) Nonvalidity of the 
Schrodinger equation concerning the "whole", O+A+E, which makes
the "state reduction process" unnecessary and unphysical, (iii)
Natural deducibility of the macroscopic irreversibility, and
(iv) Nonequivalence of MTI with any existing measurement theory,
or interpretation. Thus, within MTI, the measurement problem 
reduces basically onto the search for quantum effect, which would allow for
the local, stochastic change of Time.

\bigskip

{\bf Extended Abstract :} This is a short version of the, so-called,
Many Time Interpretation (MTI) of the quantum measurement process.
MTI involves no new hypothesis, but rather represents a new 
"re-reading' of the positive statements
of the existing quantum measurement theories. {\it Quite shortly} :
MTI relies upon the semi-epistemological analysis of the concept of
physical time (Time, or Time axis) in classical, nonrelativistic
physics. When applied to the quantum measurement situations, this 
analysis suggests abandoning the concept of unique (universal) Time 
and, in a "long run" leading to a simple interpretation
of the measurement process, {\it naturally implying} the macroscopic 
irreversibility.

We give a ("operational") {\it physical definition of Time} in classical,
nonrelativistic physics. In more subtle
analysis, our definition of Time bears significant similarities with
the {\it Aristotle-Leibnitz-Mach} approach to the same issue -
sometimes refered to as the "relational theory of time". Our definition
of Time establishes this concept as a sort of the "ordering principle"
of the classical-physics world. Actually, the definition appears
physically equivalent with the next set of the basic physical
concepts : (a) time instants and intervals, (b) existence of the 
physical laws, and (c) validity of the conservation laws.

In analysing the quantum measurement process we {\it departure from 
the widely used ensemble-in\-ter\-pre\-ta\-ti\-on}. Actually, we refer to the
{\bf single "object"} which in due course of the quantum measurement
"meets" (with some probability $W_i$) a transitional "channel" :
$\Psi \rightarrow \Psi_i$. When applied the above mentioned definition
of Time onto the "channels", one reaches the {\bf main interpretational
rule of MTI} : {\it in due course of the quantum measurement process, each 
particular "object" becomes an object of stochastic change of Time axis.
For an "object", the actual (local) Time
should be considered as objective (real) as the "macroscopic" Time 
is objective in classical physics}. 

When expressed in terms of the "macroscopic" Time, MTI
reduces onto the stochastic choice of the quantum-mechanical laws 
(still, of unknown form), each law governing the different "channel",
e.g., the i-th "channel", $\Psi \rightarrow \Psi_i$. 
This is usually refered to as "indeterminism". However, MTI {\it is 
not equivalent with this} - its somewhat naive reduction with respect 
to the "macroscopic" Time.

Still, the analysis concerning the composite systems, "object +
apparatus (O+A)", and the "object + apparatus + environment (O+A+E)",
is the full - and unavoidable - quantum-mechanical analysis.
By the use of the fact that the quantum measurement implies breaking
of some of the "conservation laws" concerning the system O+A, the role
of the environment appears both, necessary and essential in this context.
Among else, the application of the main interpretational rule of MTI
onto O+A+E implies {\it nonvalidity of the Schrodinger equation}
concerning O+A+E, and therefore the {\it process of the "state reduction"
appears both, unnecessary and unphysical}.

Besides, in an {\it axiomatic form}, MTI nontrivially extends and 
enreaches our notions concerning the "border teritory" between the 
"microscopic" and "macroscopic" parts of the physical world. Particularly,
one {\it meets amplification process as fundamental quantum process}, and
{\it obtains deducibility of the macroscopic irreversibility}.

Therefore one may say : MTI suggests abandoning the concept of universal
physical time even in nonrelativistic physics, while ascribing the
"determinism" to each local Time axis, which appears in due course of 
quantum measurement. MTI allows for overcoming the problem of "state
reduction", and in natural way provides deducibility of the macroscopic
irreversibility. Finally, the quanum measurement problem appears
reduced onto the search for a quantum effect, which would allow for
the local, stochastic change of Time.

\vfill\eject

{\bf BARE ESSENTIALS OF MTI}

\bigskip

{\bf Quick reader} should refer to the {\it Abstracts, 
Section 18, likewise to the
Outlook and Conclusion}, given on the end of this text. Below,
we give the bare essentials of MTI, and where they can be found
in the text. Especially, one may refer to the
{\it Sections 20 and 21}.

{\bf 1.} MTI proposes nonuniqueness (nonuniversality) of the
physical ("macroscopic") time in quantum, nonrelativistic physics.
({\it Cf. Sections 8-10, 12, 13, 17})

{\bf 2.} In due course of the quantum measurement process, each
singular (particular) "object" expiriences its own (local) Time (Time axis).
The choice of Time axis is stochastic. ({\it Cf. the Sections
given above})

{\bf 3.} The point "2." above is justified by considering the
composite systems, "object + apparatus", and "object + apparatus +
environment". ({\it Cf. Sections: 13-15, 17})

{\bf 4.} When reduced onto the terms which refer to the "macroscopic"
Time, MTI reproduces the results which are usually considered as
(quantum) "indeterminism". Still, the inverse is not correct (i.e.,
this reduction of MTI does not necessarily imply (lead) to MTI),
and thus can not be considered equivalent with MTI itself..

{\bf 5.} MTI implies nonvalidity of the Schrodinger (unitary) 
evolution concerning the "whole", "object + apparatus + environment".
This directly imples the "state reduction process" unnecessary and
unphysical. ({\it Cf. Sections: 14, 16})

{\bf 6.} Within an axiomatization, MTI offers a new, reacher physical
picture. Among else, one obtains possibility of simple and {\it natural
deducing of the macroscopic irreversibility}. ({\it Cf. Sections 20})

\bigskip

{\bf Instead of the "indeterminism" with regard to the standard, 
"mac\-ro\-scopic" Time, in the quantum measurement theory MTI establishes
"determinism", but with regard to the different possible (local)
Times.}

\vfill\eject

{\bf 1. INTRODUCTION}

\bigskip

This is a short version of the, so-called, Many Time Interpretation
(MTI) of the quantum measurement process.	MTI does not involve 
any new hypothesis, but rather relies upon the positive (i.e., 
interpretation free) elements of the existing measurement theories.

Here we give almost just the bare essentials, and the interesting
reader may refer to Dugi\' c 1998 for more details.

The subject will be presented "itteratively".
For the quick reader it might be convenient to refer basically to
the Sections 20-23.

Finally, we give the particular definitions of the concepts of
determinism and causality, to be used below.

By {\bf determinism} we assume {\it existence of a definite
physical state} of a system, {\it completely independent on an act
of measurement (observation)}.

By {\bf causality} we assume that, {\it given a physical state of 
a system in an instant $t_{\circ}$, the state of the system in
each later instant $t$ is known with certainty}.

\bigskip

{\bf 2. DEFINITION OF TIME IN CLASSICAL PHYSICS}

\bigskip

In this Section we give {\it an operational physical definition of
Time in classical, nonrelativistic physics}. 

This will bring us to somewhat speciffic notion on the concept of
the "macroscopic" Time as an "ordering principle" of the 
classical-physics world.

\bigskip

{\bf 2.1 Puzzling over Time}

\medskip

Usually, Time appears in physics as a {\it metaphysical concept}, which
is postulated, {\it rather than defined}. Furthermore, it is sometimes 
argued that the concept of Time, as the physicists use it, can not
be properly defined.

In our opinion, this point of view is {\it not correct}, or at least
not the only one considerable. For, from the {\it operational point
of view}, the "flow of (metaphysical) Time" in classical physics
is {\it represented ("interpreted")}
by the dynamics ("motion") of the "ideal clock" 
(cf. Appendix I). Furthermore, {\it from an operational point of view,
one may hardly ever say more about the physical Time, then it is
directly presented by the dynamics of the "ideal clock"}.
And this is the physical basis which justifies our attempt in
approaching a proper definition of the "macroscopic" Time.

\bigskip

{\bf 2.2 The "primitives" of our approach}

\medskip

The elementary observation in physics defines what we call
{\it dynamics} :
$$D = \{A, B, C, \dots \}, \eqno (1)$$

\noindent
where $A, B, C, \dots$ represent the "points" in the corresponding
physical state space, $P$, of the observed object. Usually, the
dynamics (1) is presented by some kind of "record" ("memory"),
which clearly {\it involves ordering} of the elements, $A, B, C
\dots$. And (cf. Appendix II) it is just our psychology which refers
to the "ordering" with respect to the "flow of Time" - thus implying
the circular reasoning. However (cf. Appendix II), there is no
circularity in our reasoning; in other words : we think that the
observation does not require an "instanteneous" interpretation of the
observation.

The concept of dynamics $D$ is a "primitive" of our approach and,
by definition, does not call for the more elementary ("primitive")
concepts - of course, except the concept of state space, $P$.

As another "primitive" of our approach appears the concept of
{\it "Causality"}, $C$. Note, it is not the causaliy defined in Section 1.
Actually, we {\it postulate} (cf. Appendix III for motivations)
that, given each two neighbourghing "points", e.g., $A$ and $B$ 
of the dynamics $D$, that {\it there is unique continuous "trajectory"}
in $P$, {\it connecting the two "points"}. As it was emphasized in Appendix 
III, the "Causality" is physically equivalent with the concepts of
physical law(s), i.e., with the concepts of determinism {\it and}
causality defined in Section 1 above.

\bigskip

{\bf 2.3 The concept of physical dynamics}

\medskip

From the two "primitives" of our approach, we deduce the main concept
of this Section : the concept of {\bf physical dynamics}, defined by :
$$D_p = \{A \buildrel C \over \longrightarrow B\}. \eqno (2)$$

One should note that (2) is {\it not mathematical}, but physical
expression, which should be understood : each change of physical
state $A$, to state $B$, is governed by the "Casuality" - 
which refers to existence of
unique and continuous "trajectory" governing the transition. And this
applies {\it generally} : to each physical system ("object") in classical
physics, to each particular (neighbourghing) states $A$ and $B$, 
and irrespective of physical nature of the system.

\bigskip

{\bf 2.4 Our definition of Time}

\medskip

We give the {\bf operational physical definition of Time}:
$$D_p = \{A \buildrel C \over \longrightarrow B\} :
T_M = (t^{(M)}_A, t^{(M)}_B). \eqno (3)$$

By $T_M$ we denote the "macroscpic" Time, while the pair
$(t^{(M)}_A, t^{(M)}_B)$ gives the "instants" corresponding
to the states $A$ and $B$, respectively. It is {\it not} (cf. 
Appendix IV) important if this pair is (non)unique. However,
what is - by definition - important is that : for the particular
instant $t^{(M)}_A$, the instantf $t^{(M)}_B$ is known with
certainty.

\bigskip

{\bf 2.5 Discussion}

\medskip

We do not claim that "macroscopic" Time, $T_M$, is not "real"
("objective"). For, simply, {\it it is not objective of our
considerations} ! Our objective is to obtain a {\it physically
sound definition} of Time, starting from the purely physical
data. Therefore we refer to (3) as to the {\it operational physical 
definition 
of Time}. This definition establishes physical {\it existence
of Time}, rather than it should be considered complete.

However, our approach - being concerned with the phenomenological
data - bears significant similarity with the {\it Aristotle-Leibnitz-Mach}
approach to the same issue. Sometimes, this approach is called 
"relational" theory of time (cf. Withrow 1979). Now, due to 
some shortcomings of this theory, one may wonder about the
shortcomings in our definition of Time. 

However, we do not see this appealing. Our definition of Time 
can be considered independently on the "relational" theory of time, 
on the footing established above, and  in Appendices I-IV. Still, in some
more sophisticated analyses, this might come to scope. Yet,
we think that this can be considered to be of the secondary importance
for our considerations - as we hope to become clear below.

Finally, the concept of physical dynamics - in our approach
underlying the concept of Time (although, usually
(cf. Appendix IV) it is just the 
inverse) - "brings" into the concept of Time 
the concept of "Causality". Thus one meets the concept
of Time, {\it as defined above}, as an {\it ordering principle}
in the classical-physics world; existence of the "order" follows from
postulating "Causality", which is {\it physically equivalent} with
existence of the physical laws, i.e. with the concepts of determinism 
{\it and} causality (as defined in Section 1).

\bigskip

{\bf 3. SCHRODINGER EQUATION}

\bigskip

The Schrodinger equation is the law of isolated quantum systems,
and apparently admits for introducing the concept of {\it quantum
dynamics}, in full analogy with (2) :
$$D_Q = \{\Psi_i \buildrel C \over \longrightarrow \Psi_f\}
\eqno (4)$$

\noindent
 where $\Psi_i$ is the initial state, while $\Psi_f$ is the final
 state, determined by an instant $t$ of the "macroscopic" Time, $T_M$:
 $\Psi_f \equiv \Psi_t$.
 
 Certainly, contrary to $D_p$ defined by (2), the quantum dynamics
 $D_Q$  does not appear as a result of physical
 observation. Still, $D_Q$ bears all the basic physical features 
 of $D_p$, and therefore can be {\it considered in physically
 essentially the same way, i.e., considered as a "primitive" of our 
 considerations}. 
 Particularly, the "Causality", $C$, establishes
 (postulates) existence of the unique continuous "trajectory" in the 
 Hilbert state space of the system, and is physically equivalent with
 the Schrodinger equation - $\hat U(t) \Psi_i = \Psi_t \equiv
 \Psi_f$.
 
 Certainly, the concepts of determinism and causality (cf. Section 1),
 now apply to the elements of the
 Hilbert space - as the physical state space of the
 system.
 
 Now one can note : the quantum dynamics $D_Q$ sublimates the basics
 of QM of isolated systems in the manner in which the physical
 dynamics $D_p$ sublimates the basics of classical (nonrelativistic)
 physics. Since both dynamices {\it refer to the "macroscopic" Time},
 $T_M$, one may equally state :
 $$D_Q = \{\Psi_i \buildrel C \over \longrightarrow \Psi_f\} :
 T_M = (t^{(M)}_i, t^{(M)}_f), \eqno (5)$$
 
 \noindent
 in full - certainly, physical - analogy with (3).
 
 \bigskip
 
 {\bf 4. QUANTUM STATE OF A SINGLE QUANTUM SYSTEM}
 
 \bigskip
 
 The usual, ensemble-interpretation of QM treats the ("pure")
 quantum state $\Psi$ merely as a source of informations, which
 could be provided by a proper quantum measurement procedure.
 
 An ensemble is {\it defined} as composed of the individual
 "elements" - i.e., of the single quantum systems.
 The quantum state $\Psi$ of an ensemble is assumed to be defined
 (determined) by a proper "preparation" procedure. Still, it is
 widely, although primarily implicitly, {\it assumed that each
 single system, which is an element of an ensemble in "pure" state
 $\Psi$, is also in this, "pure" state $\Psi$}. [This can be further
 elaborated, but we shall omit it here.]
 
 Bearing this statement in mind, we conclude that the dynamics (4),
 likewise the concepts of determinism and causality (cf. Section
 1), appear {\it applicable to each single system} "described"
 by the quantum dynamics (4).
 
 \bigskip
 
 {\bf 5. THE GENERAL QUANTUM MEASUREMENT SCHEME CONCERNING
 THE QUANTUM "OBJECT"}
 
 \bigskip
 
 The next scheme represents a proper generalization of the real
 quantum-measurement situations :
 $$\Psi \rightarrow \Psi_1, \quad W_1$$
 
 \noindent
 or
 
  $$\Psi \rightarrow \Psi_2, \quad W_2$$
  
  \noindent
  or ... \hfill (S1)
  
   $$\Psi \rightarrow \Psi_n, \quad W_n$$
   
   \noindent
 etc, where $\Psi$ is the initial state, while
 $\Psi_i$s represent the different final states
 ("outcomes"), while the 
 probabilities $W_i$:
 $$\sum_i W_i = 1. \eqno (6)$$

The usual, ensemble-interpretation states that (S1) can be written as :
$$\Psi \rightarrow \hat \rho = \sum W_i
\vert \Psi_i \rangle \langle \Psi_i\vert, \eqno (7)$$

\noindent
i.e., that the ensemble, initially in the "pure" state $\Psi$,
survives non-Schrodinger transition to the "mixed" state
$\hat \rho$.

It is important to note that, according to the Section 4,  the
ensemble-interpretation states that the same transition, (7), 
also refers 
to {\it each single object of quantum measurement}, i.e., that each
single "object" survives the same transition from the initial "pure"
$\Psi$, to the final "mixed" state $\hat \rho$.

\bigskip

{\bf 6. A CLUE OF MTI  1}

\bigskip

Here we propose a {\it new "re-reading" of the scheme (S1)}. Our proposal
should not be understood to claim that the usual, the ensemble-interpretation,
is not correct, but just that it is not the only one possible.

What we suggest is to distinguish the different transitional "channels"
in
(S1) : the first "channel" refers to the transition $\Psi 
\rightarrow \Psi_1$, the second "channel" to the transition
$\Psi \rightarrow \Psi_2$, etc.

Now, and this is the point to be strongly emphasized, {\it as regards the
single "objects"}, one may note that : {\it each single "object" has a
choice between the different transitional "channels"}. This choice is
governed by the corresponding probability distribution, $\{W_i\}$, which
must be tested on an ensemble. I.e., the choice of the "channel" should be 
considered to be stochastic.

Therefore, {\it we propose} to consider the transitional "channels",
e.g., the i-th one :
$$\Psi \rightarrow \Psi_i, \eqno (8)$$

\noindent
{\bf as the real physical process for the actual  single
"object"}. Therefore, in this context,
the physical place of the probabilities $W_i$ is yet to be determined.

\bigskip

{\bf 7. COMMENTARY}

\bigskip

Needless to say, the transition due to the i-th "channel" appears as a
quantum analogue of the classical dynamics (1) which defines the
dynamics of the quantum measurement process :
$$D_i = \{\Psi, \Psi_i\}. \eqno (9)$$

\noindent
Besides, after the measurement has ceased (relative to the "macroscopic"
Time $T_M$), the evolution of the object is governed by the Schrodinger 
law.

Here we make an ansatz concerning the "channels" : actually, we assume
existence of the physical law which should govern the transition due to
the given "channel" :
$$\hat U_i \Psi = c_i \Psi_i, \eqno (10)$$

\noindent
where $c_i$ is a "complex number", 
and which is a consequence of, in general, that the
"evolution operator" $\hat U_i$ needs not to be unitary. Certainly,
there is the time dependence $\hat U_i = \hat U_i(t^{M)})$, which is
for the simplicity omitted above. 

The presumption concerning {\it existence} of $\hat U_i$ follows from
our the main assumption : {\it that the transitional "channels" refer to
the real physical processes}. Then. certainly, there must exist the
law Eq. (10) (and also
its the differential form). [It is obvious that we
just state existence, but not the mathematical form of the law (10).]

In the usual {\it ensemble-interpretation} the existence of $\hat U_i$s
would be interpreted as a formal expression of {\it "indeterminism"}.
However, we offer a new point of view, which is the basis of MTI.

\bigskip

{\bf 8. A CLUE OF MTI 2}

\bigskip

The very existence of $\hat U_i$s admits for introducing the concept
of {\it quantum dynamics in the context of the quantum measurement process},
by :
$$D_{Qi} = \{\Psi \buildrel C_i \over \longrightarrow \Psi_i\},
\eqno (11)$$

\noindent
and which should be read (understood) in full analogy with (3) and
(4), bearing in mind that the index "i" in $C_i$ refers to the i-th
operator $\hat U_i$ - cf. (10). [Certainly, we assume that the "Causality"
$C_i$ refers to unique and continuous "trajectory" in the Hilbert space,
but this is not substantial assumption here - cf. Appendix V.]

Note that all the physical dynamices, (3), (4) and the stochastic one,
(11), {\it have the same physical contents : they all refer to the single
"objects", bearing determinism and causality}
(with regard to the Hilbert state-space). This is what
gives us right to make another, {\it substantial} step in our
interpretation, i.e., to define the {\bf different Time axes} :
$$D_{Qi} = \{\Psi \buildrel C_i \over \longrightarrow \Psi_i\} :
T_i = (t^{(i)}, t^{(i)}_i), \eqno (12)$$

\noindent
bearing some redundancy in the indices.

The expression (12) is the very form of the {\bf main interpretational
rule of MTI} : {\it In due course of the measurement process, each single 
"object" becomes an object of stochastic - with probability $W_i$ -
change of Time, i.e., of the choice of the Time axis, $T_i$. 
Each Time axis $T_i$ is as real
(objective) for the actual "object", as the "macroscopic" Time axis, 
$T_M$, is real
in the classical physics}. [Therefore, the physical "origin" of the 
probabilities $W_i$ is yet to be determined, by determining an effect
which should provide the stochastic change of Time.]

\bigskip

{\bf 9. A CLUE OF MTI 3}

\bigskip

We propose the Time axes $\{T_i\}$ to be considered seriously. That is,
that the "macroscopic" Time, $T_M$, should be considered just as an
{\it example of physical Time}. And : {\it we have abandoned the concept
of unique (universal) Timef even in nonrelativistic physics, but have
ascribed the determinism and causality  to each Time
axis}. In so far as we can see, this is in no contradiction with the present
state of art in quantum measurement theories.

Intuitively, MTI suggests introducing the sudden, stochastic choice of
{\it local physical worlds}, each of which is established and "ordered"
in the same way (qualitatively), as is the case with the classical
("macroscopic") physical world, which is "ordered" by $T_M$.
 Certainly, the macroscopic objects are
the objects of the macroscopic Time, and therefore of the macroscopic
physics laws. Similarly, an "object" expiriences its own local Time axis 
(world), which defines its "own" quantum laws, which can be thought
about (cf. Section 11), but not directly expirienced by the macroscopic
bodies. This further means that the macroscopic clocks can only and 
exclusively measure the "macroscopic" Time. What then one can admit as
a "clock" for the "microscopic" Times ? {\bf The answer is : the dynamics 
of the "objects", $D_{Qi}$, itself}. I.e., the "object" itself represents
(for definition cf. Appendix 1) an ideal clock for the corresponding
Time axis. We now hope to be clear that what we should do
at this point, is to make some connections between the instants and
intervals of the "macroscopic", and the local ("microscopic") Times.

\bigskip

{\bf 10. ONE TIME OR MANY TIMES ?}

\bigskip

One may wonder if the interpretation in terms of the different Time axes
is really necessary. There are a few answers with respect to the different,
implicit {\it aspects to this dilemma}.

{\it First}, this interpretation is not necessary - otherwise, the 
standard one
would never come to scope. What we claim is that MTI is not
forbidden.

{\it Second}, {\it we use the term} Time {\it as 
physically essentially equivalent
with the concept of physical/quantum dynamics}. [This is justified by the
definition of Time in Section 2, and the fact that all the basic features
of the physical dynamics, (2), appear in (4) and (11) - as it is 
distinguished
in Section 8]. However, if one would claim the opposite - i.e., that
(5) and (12) can not be considered as a recognition of the different {\it
Times} - this would mean that, physically, we know much more about Time
than it is stated by the definition (3); for our point of view see
Appendix I.

{\it Finally}, one may note that the operators $\hat U_i$ (cf. Section 7)
appear physically sufficient (e.g., as a formal expressions 
of the standard
"indeterminism"), therefore offering no new contents. We note that
{\it MTI reduces itself onto this picture, but with regard to the
one and unique (universal), "macroscopic" Time}; that is,
MTI appears {\it reducible} onto the set of the given operators, 
$\hat U_i$, of Section 7.
However,
and this is the point to be emphasized, {\bf this reduction of MTI does not
appear equivalent with MTI itself}. And this is going to be proved below
(cf. Secion 12).

Therefore, it is admitable to {\bf deal with nonunique Time in the context
of the quantum measurement theory}.

\vfill\eject

{\bf 11. A CLUE OF MTI 4 : SOME FORMAL EXPRESSIONS}

\bigskip

If each Time axis, $T_i$, should be considered physically equal with 
the "mac\-ro\-sco\-pic" Time, $T_M$, then {\it everything that 
can be physically
stated in terms of $T_M$, should be
expressible in terms of the instants (intervals) of each local Time,}
$T_i$. Instead of being exhaustive in this respect, let us refer 
to the operators $\hat U_i$ of Section 7.

According to the rule of Section 8, 
one should refer to an operator $\hat U^{(i)}$,
which should be seen as the "evolution operator" concerning the Time
axis $T_i$ - i.e., concerning the i-the "Causality", $C_i$. Certainly,
this operator defines the quantum law, which refers to the i-th
"channel", and can be written as :
$$\hat U^{(i)} \equiv \hat U^{(i)}(t^{(i)}). \eqno (12)$$

When compared to Eq. (10), the expression (12) leads to equality :
$$\hat U_i(t^{(M)}) = \hat U^{(i)}(t^{(i)}), \eqno (13)$$

\noindent
which implies (as it is required in Section 10) {\it connectability}
of the instants and intervals of the Time axes, and particularly :
$$dt^{(i)} = g_i(t^{(M)}) dt^{(M)}, \eqno (14)$$

\noindent
with obvious meaning, while generally: $g_i(t^{(M)}) \neq 1$.

However, it is interesting to note that (14) has an important, direct
consequence. Actually, if one may write :
$${d\hat A \over dt^{(M)}} = 0, \eqno (15)$$

\noindent
then (14) directly implies :
$${d\hat A \over dt^{(i)}} = 0. \eqno (16)$$

Therefore, the {\it requirement of the full physical equality of the Time
axes is fulfilled}: 
{\it each Time axis has its own "instants" and "intervals", defining the
corresponding quantum law for the actual  "object", 
and alowing for the general validity of the conservation laws}.

[Finally, we offer a specific speculation, which seems both, physically
plausible and welcome, likewise simplifying our interpretation. Actually,
it seem un-forbidden to assume that {\it there is unique (quantum law)}
for both, {\it isolated and open quantum systems}, which is represented by
some operator (of still unknown characteristics), $\hat U^{(u)}$. 
The universality
means that one may state:
$$\hat U^{(u)}(T_i) = \hat U^{(i)}(t^{(i)}),$$

\noindent
including
$$\hat U^{(u)}(T_M) = \hat U(t^{(M)}),$$

\noindent
where $\hat U(t^{(M)})$ is the unitary Schrodinger's operator.
Above, $\hat U^{(u)}(T_i)$ should be considered as a{\it 
"re\-pre\-sen\-ta\-ti\-on}
of $\hat U^{(u)}$ with respect to $T_i$

Certainly, to make sense, this hypothesis should firstly provide
us with more precise physical meaning of the above "representation" 
given above.
Still, it significantly simplifies MTI, by assuming existence of
unique quantum law at "all [physical] levels" - cf. Penrose 1994
(p. 308).

Fortunatelly, if this hypothesis would finally prove unjustified,
the MTI itself would not "suffer" from this, at all.]

\bigskip

{\bf 12. THE ISOLATED COMPOSITE SYSTEM "OBJECT PLUS APPARATUS"}

\bigskip

Reducibility of MTI onto the interpretation given in Section 7
is probably obvious (cf. Section 10). Here we show that the 
inverse does not prove correct.

The real object of the von Neumann's theory is not the "object" alone,
but the composite system, "object + apparatus (O+A)". In the original 
theory, the system O+A is considered {\it isolated} - meaning that 
one must consider the Schrodinger equation valid for O+A.

As regards O+A the Scheme (S1) appears extended as:
$$\Psi \chi \rightarrow \Psi_1 \chi_1, \quad W_1$$

\noindent
or
$$\Psi \chi \rightarrow \Psi_2 \chi_2, \quad W_2$$

\noindent
or ... \hfill (S2)
$$\Psi \chi \rightarrow \Psi_n \chi_n, \quad W_n$$

\noindent
etc. By $\Psi$s we denote the states of the "object", and by $\chi$s
the states of the "apparatus". 

Again, one can recognize the "channels" in (S2), and therefore, in
analogy with (10), may introduce the operators $\hat V_i$ (instead of
$\hat U_i$s of Section 7).

{\bf However, the set $\{\hat V_i\}$ does not 
necessarily lead to MTI !}

This can be seen as follows. The operators $\hat V_i$ are defined by:
$$\hat V_i \Psi \chi = c_i \Psi_i \chi_i, \eqno (17)$$

\noindent
but {\it which could be in accordance with the Schrodinger
equation}:
$$\hat U \Psi \chi = \sum_i d_i \Psi_i \chi_i, \eqno (18)$$

\noindent
where $\hat U$ is the unitary operator of the Schredinger law.

As it can be easily seen, this can be fulfilled if one
may state $c_i = d_i, \forall{i}$, and :
$$\hat U = \sum \hat V_i. \eqno (19)$$

Therefore, as it was told above (cf. also Section 10), 
the very existence of the
operators $\hat V_i$ - in analogy with (10) - {\it does
not necessarily lead to MTI, but} - under the condition (19) -
{\it leads to the von Neumann's theory}.

Needless to say, in the von Neumann's theory, the expression
(19) - cf. Section 4 - refers to both, an ensemble of pairs, 
O+A, and to each single pair. Therefore, as the 
{\bf real physical process}, in this theory, appears the 
Schrodinger equation - and how otherwise could be - "in" the
"macroscopic" Time? Then, {\bf the operators} $\hat V_i$
{\bf appear artificial, without any physical meaning}, and
can be eventually used as a mathematical tool. And this is
exactly from what suffers the operators $\hat U_i$ of Section 7, 
in the same context.

Relative to this, MTI {\bf states exactly oposite} : in terms of this
Section, MTI states that each operator (i.e., the corresponding
"channel") $\hat V_i$ refers to the {\bf real physical process}. Then, if 
(19) would appear admitable, the unitary operator, $\hat U$
- cf. l.h.s. of (19) - appears {\bf unphysical, artificial},
and can be only used for mathematical convenience. Certainly, in full
analogy one gives an answer to the question 
raised in Section 10 concerning whether
the very existence of the operators $\hat U_i$ (of Section 7) can be seen
equivalent with MTI, thus justifying told therein, and in the 
begining of this Section.

\bigskip

{\bf 13. THE OPEN SYSTEM  O+A}

\bigskip

The previous analysis refers to the original von Neumann's
theory. However, the system O+A is really an {\bf open system},
which is due to the openess of the "apparatus".

The openess of the "apparatus" is not just a fruitfull hypothesis
in the modern decoherence theory. It is a statement that also follows
from the more elaborated von Neumann's theory. Actually, as it was 
shown by Araki and Yanase 1960 and Yanase 1961, the {\bf quantum 
measurement process breaks some conservation laws concerning the
system O+A}. Certainly, this implies existence of the "environment" (E),
which should provide validity of the conservation laws - certainly,
now for the "whole", O+A+E. Furthermore, as it was shown by Zurek 
1983 (p. 93), this task can be done correctly {\it if} there is a
part E'' of the environment which is not in correlation with the
"rest", O+A+E', E'-the correlated part of the environment ; E = E' + E''.
In other words, the part E'' is {\it "here just to pay for the balance"}.

Therefore, the composite system O+A {\it should be considered in analogy
with the object O} ; i.e., the scheme (S2) should be considered in analogy
with the scheme (S1).

However, if the states of the "apparatus" in (S2), $\chi$s, should be
interpreted as {\it mutually macroscopically distinguishable}, then
the direct interpretation in terms of many Time axes - which should also
refer to the macroscopic variables - would appear incorrect. Actually,
one generally deals with the macroscopic variables as the objects of
the "macroscopic" Time, bearing {\it classical reality} (determinism, causality
and locality).

However, there is another positive statement of the measurement theories, 
which {\it makes MTI sound even in this context} ; this is the {\bf amplification 
process}. Actually, according to this, the "apparatus" consists in the
two parts, the "microscopic" one, A', and the "macroscopic" one, A'' ; 
A = A' + A''. Then the measurement has the two stages : 
$$\Psi_O \phi_{A'} \Phi_{A''} \buildrel T_i \over \longrightarrow
\Psi_{Oi} \phi_{A'i} \Phi_{A''} \buildrel amplific. \over \longrightarrow 
\Psi_{Oi} \phi_{A'i} \Phi_{A''i}. \eqno (20)$$

The first stage (cf. the 1st arrow from the left) establishes the 
correlations between O and A'. Since both systems, O and A', are the 
"microscopic" ones, one may {\bf apply the main interpretational rule of
MTI onto O + A'}; note: $\chi_A \equiv \phi_{A'} \Phi_{A''}$,
$\chi_A$ appearing in (17). 
The second stage (cf. the second arrow) is the
{\bf amplification process}. It consists in transfering the information
"contained" in A', to the "macroscopic" part, A''. This transfer can be
complex, and we shall postpone its discussion until subsection 20.6.
Here we just want to note that the second stage {\it does not admit for
applying the stochastic change of Time axis, and which is due to the
macroscopicity of} A''. [Note : {\bf everything still comes by definition,
but will be stroingly justified in Section 20.}] It is essential to note{
that the operators $\hat V_i$ in Section 12 now appear as just {\it effective
transfornations}, i.e. (contrary to $\hat U_i$s of Section 7) not defining
the Time axes, $T_i$. Rather, $\hat V_i$s are defined by (17), i.e., by
$\hat V_i \Psi_O \phi_{A'} \Phi_{A''} = \Psi_{Oi} \phi_{A'i} \Phi_{A''i}$,
without stating the details concerning the involved, local
Times, which are distinguished in (20).

It can not be overemphasized that each system, O, A', and A'' has its
own {"pure"} state in each stage of the process, and allows for the complete
application of MTI in the first stage of the process (20).

Therefore, we conclude that applicability of MTI concerning O+A
{\it requires} existence of the amplification process.

\bigskip

{\bf 14. THE "WHOLE"  O+A+E}

\bigskip

Certainly, the complete analysis refers to the "whole", O+A+E =
O + (A'+A'') + (E'+E''). For the simpolicity we shall omit the later, 
i.e., the precise "decomposition" of the apparatus and of the environment,
bearing im mind that, in accordance to the previous Section, the "correlated"
part of the environment, E', should consist in the two parts, the
"microscopic" one, $E'_1$ and the "macroscopic" one, $E''_2$. Further,
we shall also assume the underlying, local amplifaction
processes concerning both, A, and E'. 

Now the contents of the previous Sections can be summarized as follows :
$$\Psi_{O} \chi_{A} \lambda_{E'} \kappa_{E''} \rightarrow
\Psi_{Oi} \chi_{Ai} \lambda_{E'} \kappa_{E''} \rightarrow
\Psi_{Oi} \chi_{Ai} \lambda_{E'i} \kappa_{E''}, 
\quad W_i, \eqno (21)$$

\noindent
without the details (the "parts", and amplifications), 
and with obvious notation.

Let us note : The evolution of the"whole" clearly {\it distinguishes the two
parts of the "whole"}. The one part, O+A+E', bears the correlations. The second
one, E'', does not. Certainly, the evolution of the first part is complex,
involving the different stages concerning each "arrow" in (21). Further, 
it bears existence of the local Time
axes for each single "whole" - i.e., in each single run of the 
measurement. On the other side, there is the uncorrelated part E'' of
the environment, which is "here just to pay for the balance". The quantum
state of E'' does not significantly change - as it is presented above
(cf. Zurek 1983, p. 93).

Now, to this end, the main interpretational rule of MTI states: 
{\bf a single "whole"
O+A+E can not be considered to evolve according to unique Time axis}.
Furthermore, due to the complexity of the "whole", {\it each single "whole"
has its own set of the local Time axes}, each being
governed by the corresponding probability.

\bigskip

{\bf 15. QUANTUM STATE OF THE ENSEMBLE}

\bigskip

The expression (21) refers to each single "whole". Due to MTI, each 
single "whole" has its own {\it final} ("pure") state :
$$\Phi_{fO+A+E} = \Psi_{Oi} \chi_{Ai} \lambda_{E'i} \kappa_{E''},
\eqno (22)$$

\noindent
with some probability, $W_i = \lim_{N \to \infty} N_i/N$, where
$N_i$ represents the number of appearance of 
the i-th outcome, while $N$ represents
the total number of the outcomes. 

Now, {\bf by definition}, the state of the ensemble of $N$ elements
(here : single "wholes") reads :
$$\hat \rho_{fO+A+E} = \sum_i W_i \hat P_{iO+A+E} \otimes
\vert \kappa_{E''}\rangle \langle \kappa_{E''}\vert, \eqno (23)$$

\noindent
where $\hat P_{iO+A+E} \equiv \vert \Psi_{Oi}\rangle \langle 
\Psi_{Oi} \vert  \otimes
\vert \chi_{Ai}\rangle \langle \chi_{Ai}\vert \otimes 
\vert \lambda_{E'i}\rangle \langle \lambda_{E'i}\vert$.

It is essential to note : (i) The state (23) is exactly the one
defined by the "projection postulate", (ii) {\bf Ensemble} of each 
subsystem, O, A,
E (and their parts distinguished above) has unique "mixed" state,
e.g.:
$$\hat \rho_O = Tr_{A+E} \hat \rho_{fO+A+E} =
\sum_i W_i \vert \Psi_{Oi}\rangle \langle 
\Psi_{Oi} \vert , \eqno (24)$$

\noindent
and analogously for A, E, and their "parts".

And at this point we meet the two distinctions between MTI and the  usual 
ensemble-interpretation. The one will be presented in Section 17, while 
there is another one : The "mixture" (24) {\it really is} the "improper
mixture" (cf. d'Es\-pag\-nat 1976), but it refers exclusively to the {\it ensemble
point of view}. Here, in MTI, the objective is the single system
(O, O+A, or O+A+E). Now, given the state of each single system, the state
of the ensemble directly follows. But, and this is the point, there
is no any doubt concening the quantum state of any single system in MTI.
I.e., {\it the problem of the "improper mixtures" disappears 
in MTI, due to being concerned with the single systems}.

\bigskip

{\bf 16. IRELEVANCE OF THE "STATE REDUCTION PROCESS"}

\bigskip

Now we are prepared to give probably the very central implication of
MTI: physical irrelevance (i.e., unphysical character) of the von
Neumann's "state reduction process".

As it was emphasized in Section 14, {\it the "whole" O+A+E admits for 
preserving the conservation laws} (cf. the refs. therein), {\it but can not
be considered as an object of unique Time}. Furthermore, {\it there is a
collection of the local Time axes for each single "whole"}. Needless
to say, this physical situation can not be considered equivalent with
existence of the unique, "macroscopic" Time, as is the case in the
von Neumann's theory, in which the whole ensemble, likewise each
particular (single) "whole", is an object of the "macroscopic" Time.

The unique Time in the von Neumann's theory defines the
Schrodinger law as the quantum law for both, single "whole", likewise
for an ensemble of the "wholes". The  Schrodinger equation bears linearity
and establishes the {\it correlations}:
$$\hat U \Psi_O \chi_A \Phi_E = \sum c_i \Psi_{Oi} \chi_{Ai} \Phi_{Ei},
\eqno (25)$$

\noindent
and therefore {\bf calls for the "reduction process"}. 

However, in MTI, {\it the final state (23) is obtained by the main interpretational
rule of MTI}, bearing the different sets of the local Time axes
for each single "whole". Certainly (cf. also Section 17 for some details),
this makes the Schrodinger evolution {\bf unphysical}, and therefore, the
"state reduction process" {\bf unnecessary and unphysical}.

\bigskip

{\bf 17. THE ISOLATED O+A+E}

\bigskip

Now one may state a question raised in Section 10 (and essentially
answered in section 12), but with regard to the "whole", O+A+E.

Actually, the Scheme (S2) should be extended: everywhere instead of 
$\Psi \chi$, should be put $\Psi \chi \Phi$, where $\Phi$ refers to
the state of the environment. Again, there appear the "channels" of
the type $\Psi \chi \Phi \rightarrow \Psi_i \chi_i \Phi_i$, which are 
governed by some (also - cf. Section 13 - effective)
evolution operators $\hat W_i$ - instead of the operators
$\hat U_i$ (cf. Section 7) for the "object" alone, and (cf. Section 13)
the effective operators $\hat V_i$ for the 
composite system, O+A.

{\it Again, the very existence of the operators $\hat W_i$ does not appear 
equivalent with MTI, in the exactly the same way as it was shown in
Section 12}. And this is another point at which one meets significant
distinction between MTI and the von Neumann's theory. Let us briefly 
repeat what was told in Section 12. but here bearing in mind the "whole".

The existence of $\hat W$s (even if these were not effective operators)
is {\bf not} in contradiction with the
von Neumann's theory, while MTI {\bf is} in contradiciton with the 
von Neumann's theory. Actually, the Schrodinger law appears to be the real
physical process in the von Neumann's theory, while then the operators
$\hat W_i$ prove just to be there for mathematical convenience, bearing no
element of reality.

On the contrary, in MTI one refers the operators $\hat W_i$ to 
the {\bf real physical processes concerning the single "wholes"},
while the Schrodinger equation appears unphysical.

Certainly, the expression $\hat W_i \equiv \hat W_i(t^{(M)})$ represents
reduction of MTI onto the "macroscopic"
Time - i.e., presenting $\hat W$s in terms of the "macroscopic" instants.
As it is stated above: this reduction {\bf allows also for the von Neumann's
interpretation}, this reduction of MTI
can not be considered equivalent with MTI itself,
and {\bf necessarily calls for the Many Times} - cf. on the end of Section
10 - {\it instead of only one Time}.

\bigskip

{\bf 18. TENTATIVE CONCLUSION}

\bigskip

We have consistently applied the main interpretation rule of MTI to the
composite systems, O+A, and O+A+E, with no new hypothesis.

The application leads to : (i) Necessity of the amplification process
in the measurement process, (ii) Obtaining the "mixed" state of each
subsystem in exactly the same form as stated by the "projection
postulate", (iii) Unphysical character of the Schrodinger equation
concerning the "whole", O+A+E, and therefore unphysical character of 
the"state reduction process", and (iv) Nonequivalence of MTI with the
von Neumann's theory, or with the reduction of MTI onto the "macroscopic"
Time.

Finally, each subsystem of a single "whole" has a definite quantum state
- which is the (quantum) {\bf determinism} - (cf. Section 1 for  
definition).
On the other side, each local Time refers to the unique transition,
i.e., to the unique final state - which is the (quantum) {\bf causality}.
Therefore, the analysis concerning the composite systems justifies the
analysis, which refers to the "object" alone. Finally, MTI {\bf reduces 
the quantum
measurement process} onto the search for the quantum effect, which would
allow for the local, stochastic change of Time axis; this gives 
an "interpretation" of the probabilities, $W_i$, of the
schemes (S1) and (S2).

\bigskip

{\bf 19. SOME CRITICAL REMARKS}

\bigskip

One may note that the final state $\hat \rho_{fO+A+E}$, Eq. (23), comes
from MTI {\it by definition}. On the other side, one would expect that the 
process of change of Time should bear a deeper physical foundation. Thus
one may wonder if the rules of MTI might bear some generality, and 
probably wider applicability.

On the other side, a "solution by definition" virtually bears dangeour.
E.g., if understood literally, by this one 
might "explain everything", by simple
asserting that "there is a such Time axis"...

All this actually points to a need for {\it axiomatization} of MTI.

\bigskip

{\bf 20. AXIOMATIZATION OF MTI. A NEW PHYSICAL PICTURE}

\bigskip

Here we shall adopt the plausible definition of the 
"macroscopic" ("classical") systems which is usual
in, e.g., modern decoherence theory: by the "macroscopic"
systems we shall assume the many-particle systems which are in
unavoidable interactions with their environments. [Certainly, the
"microscopic" ("quantum") systems are those which can be,
at least in principle, considered isolated.] Further, we assume
existence of the macroscopic systems as  phenomenological data,
without dealing with - in epistemological terms - the question of
{\it becoming} of the data; still, this does not mean that the "data" 
can not be eventually deduced within MTI.

Finally, we assume that each measurement-like interaction
defines a set of
possible Time axes, while the very possibility of changing Time axis
{\bf requires} existence of a part E'' of environment, which
"should pay for the ballance" (cf. Sections 13 and 14). An isolated
quantum system is certainly an object of "macroscopic" Time, $T_M$.

Bearing these assumptions in mind, we shall formulate the particular
propositions, and investigate some of their basic implications and
predictions.

\bigskip

{\bf 20.1 The propositions of MTI}

\medskip

{\bf (P1)} {\it Each physical system should be considered as  an object of
one and only one Time axis. Each Time axis defines "its own" physical
laws.}

{\bf (P2)} {\it Each two, mutually relatively strongly interacting systems,
have a common Time axis.}

Let us briefly clarify the two propositions.

First, {\bf (P1)} establishes that a physical system {\it evolves
due to a definite Time axis}. Therefore, the set
of the alternative Time axes appears as a sort of the classical
stochastic variable - there are no the "coherent mixtures" of the
different Time axes. Each
physical situation is governed by a physical law, but the 
{\it form of the law
is determined by the actual Time axis}.

As regards {\bf (P2)}, one should note that a system can interact
with many physical systems; i.e., there might be a numerous
set of interaction Hamiltonians, which depend upon the observables
of the actual system. Yet, only some of these Hamiltonians would
in general appear "effective", i.e., non-negligible. This defines
what was called in {\bf (P2)} as " relatively strong interaction".
Certainly, {\bf (P2)} states that not necessarily all the interacting
systems should be considered to have a Time axis in common; just some
of them, which are "relatively strongly" interacting. Needless to say,
this provides {\bf locality of the Time axes}. [Note : this locality
does not have much in common with the "locality" in the context of the
EPR paradox - to be discussed in subsection 20.9.]

Therefore, the above propositions provides us with {\it determinism,
causality and locality} concerning the isolated quantum systems,
likewise the objects of the quantum measurement processes.

\bigskip

{\bf 20.2 Existence of the "macroscopic" Time}

\medskip

This is the first implication (prediction) of the proposition {\bf
(P2)}. 

As we told above, we assume existence of the macroscopic systems,
$S_1, \quad S_2$, $\dots S_n$. By definition, each macroscopic 
system is an open 
system, i.e., the complete picture concerning the macroscopic systems
is the next one : $(S_1+E_1), \quad (S_2+E_2), \dots (S_n+E_n)$,
where $E$s denote the corresponding environments.

This situation can be re-written as $S + E$, where
$S = \sum_i S_i$, and $E = \sum_i E_i$; let us refer to $S$ and $E$
as to the macroscopic part of the Universe, and its environment, 
respectively. 

However, it is the very nature of the macroscopic systems that they are,
pairwise, in interaction, for instance:
$S_1-S_2$, $S_2-S_3, \dots$, thus making the
"chain" of interacting systems. 
Let us assume that
all these interactions are relatively strong. 

Then there is an {\it immediate consequence of the proposition}
{\bf (P2)} : as regards the above "chain", the proposition {\bf (P2)}
implies that {\it the "chain" should be considered as an object of the
same Time} - which, {\it by definition}, is the "macroscopic" Time, measured
by the macroscopic clocks.

Certainly, the two parts of the "chain", $S_1$ and $S_n$, although
mutually only weakly interacting, are both objects of the "macroscopic"
Time. But this is {\it not} due to its 20.3
{\it mutual} interaction, but due to
the {\it "chain"-character} of the system $S$. And everything directly
applies to the environment $E$, likewise to S+E

\bigskip

{\bf 20.3 The "structure" of the "macroscopic world"}

\medskip

Let us refer to the above defined system $S+E$ as to the 
"macroscopic world".

Above, we have assumed that each macroscopic system, $S_i$, could be
considered to be "solid". However, each system has its own "structure"
- divisibility into the subsystems. And not all the subsystems would
be "relatively strongly" interacting with the rest of the macroscopic
body. This highly plausible notion further leads to making another picture 
of "structure" of macroscopic body (further : body) : there
are some parts ("microscopic" subsystems), 
which are relatively weakly interacting
with the mu\-tu\-a\-lly-re\-la\-ti\-ve\-ly-strong\-ly-in\-te\-rac\-ting
pieces ("macroscopic" subsystems) of the body. Now, the later 
is a part of the body
which {\it directly refers} to the "macroscopic" Time. Still, the choice
of the Time axis {\it depends on the locally strongest interaction}.
Therefore, in so far as the "microscopic" part of the body is in relatively
strong interaction with the "macroscopic" one, it is also an object of 
$T_M$.

{\bf Note:} what is the "microscopic", and what the "macroscopic"
part(s) of a macroscopic body is here undefined. However, we believe that
in each particular situation, this can be properly defined without any
serious obstacles.

\bigskip

{\bf 20.4 "Embedding" of the macroscopic bodies in
$T_M$. Origin og the macroscopic quantum fluctuations}

\medskip

One may wonder if the macroscopic bodies can be considered as the
objects of stochastic change of Time. But the answer is : {\bf NO},
despite the fact that the macroscopic bodies are (cf. above) defined
as the open quantum systems. This is simply because that {\it there is
no such a big environment} (more precisely: the part E''), {\it which should
"pay for the balance"}.

However, strictly speaking, this refers to the "macroscopic" subsystems
of the macroscopic bodies : these {\it are "embedded" in} $T_M$. Still, the
very existence of the "microscopic" pieces defined above, allow for
the next considerations : Let us consider the two bodies, $S_1$ and 
$S_2$; the "microscopic" pieces are $\sigma_1$ and $\sigma_2$, respectively.
The interaction between $\sigma_1$ and $\sigma_2$ might exced all
the other interactions: $\sigma_1-S'_1$,
$\sigma_1-S'_2$, $\sigma_2-S'_1$, $\sigma_2-S'_2$ ; here : 
$S'_i = S_i/ \sigma_i, i =1, 2$.

Now, {\bf (P2)} applies to the (locally, relatively strong) interaction
$\sigma_1-\sigma_2$. If the two systems are both sufficiently 
"microscopic" (cf. counterexample on the begining of this subsection),
one meets the possibility of the (local) change of Time axis,
concerning the pair $\sigma_1 + \sigma_2$. Now the amplification
from $\sigma_{1,2}$ to $S_{1,2}$ provides us with
another prediction of MTI which can be eventually
recognized as the origin of the "macroscopic quantum fluctuations".

OK, but what when (measured by macroscopic clock)  
this interaction would become "weak" ? The answer will be given in 
subsection 20.6. Let us just remind that
then the interactions $\sigma_1-S'_1$, and $\sigma_2-S'_2$, become 
relatively strong interactions.

\bigskip

{\bf 20.5 The Hamiltonian equations}

\medskip

As it was told in introduction of this Section and in subsections 20.1
and 20.2, both, the isolated "microscopic", and the open "macroscopic"
systems are the objects of the "macroscopic" Time, $T_M$. Due to
{\bf (P1)}, there is another direct "prediction".

Let us for simplicity be concerned with the {\it mechanical systems}. 
Then {\bf (P1)} implies - cf. the second part of {\bf (P1)} -
that both kinds of systems (classical and quantum)
{\it should be governed by the same kind
of physical law(s)}. And this is exactly the case - e.g., the Hamiltonian
equations. [At this point one meets the limitations of MTI : 
we give the
formal correspondence between the classical and quantum systems,
without entering the subtleties concerning the corresponding state spaces,
and the deeper questions concerning the
physical interpretation of the "wave function".]

\bigskip

{\bf 20.6 The amplification process}

\medskip

In Section 13 we have distinguished the process of amplification as
 a substantial stage in the measurement process. Let us put this with
 some detail.
 
The situation we are concerned with  is the next one: a microscopic
"object" is in the measurement-like interaction with a macroscopic 
apparatus, A. The result of this interaction should fit (S2).

Before the interaction, the "microscopic" part A'of the apparatus
is in relatively
strong interaction with the "rest", $A''(=A/A')$. Due to {\bf (P2)},
then A' is also an
object of $T_M$. When the interaction between
O and A' becomes dominant, the first stage of the
measurement process (cf. (20)) can proced. Note: then the interaction
between A' and A'' becomes negligible, but not exactly zero; i.e., 
A' {\it is an open system, while A'' is not (cf. 20.4) an object of change of
Time}, for the two reasons: (a) it is weakly interacting with A',
and (b) A'' is embedded in $T_M$.

Due to the openess of A', the composite system O+A' is also an open
system, and, according to {\bf (P2)}, might become an object of the
stochastic change of Time :
$$\Psi_O \phi_{A'} \buildrel T_i \over \longrightarrow
\Psi_{Oi} \phi_{A'i}. \eqno (26)$$

However, after some time (measured by macroscopic clock), the
interaction between O and A' will cease to be dominant. Then the
dynamics of O, according to MTI, becomes governed by the Schrodinger
law. 

On the other side, however, the interaction of A' with A'' {\it again
becomes dominant}, and this is the begining of the second stage
(cf. the second arrow in (20)) of the measurement: {\bf the aplification
process}. Then the information "stored" in A', transfers to A'', thus
giving rise to :
$$\phi_{A'i} \Phi_{A''} \buildrel amplific. \over \longrightarrow
\phi_{A'i} \Phi_{A''i}. \eqno (27)$$

There are a few scenarios for (27). They can be classified relative
to the next criteria : whether one considers QM universally valid or not, 
and concerning the "structure" of A'', which might provide the different
stages in the amplification itself. Here we shall refer to 
probably the simplest situation : universally valid QM, without
the "stages" in amplification.

Now, MTI implies: {\bf A'' is an object of $T_M$}. Then the direct
amplification (interaction of A' with A''), due to "embedding" of A''
in $T_M$,
would mean {\bf "glueing" of A' to $T_M$}; i.e., A' survives
another change of Time, and the {\it final Time (with certainty) is}
$T_M$. Now, in the 
context of the {\it universally valid QM}, MTI implies that
the physical law is the Schrodinger
law, i.e. :
$$\phi_{A'i} \Phi_{A''i} = \hat U \phi_{A'i} \Phi_{A''}, \eqno (28)$$

\noindent
where $\hat U$ is the unitary Schrodinger operator concerning
the composite system A'+A''. (Needless to say, the expression (28)
calls for the specific interaction Hamiltonians, which can be found in
Dugi\' c 1996/97.)

[Finally, the above scenario refers to an ideal quantum measurement,
while the "steps" in amplification might involve the slight changes in
the state of A' - non-ideal measurement.]

\bigskip

{\bf 20.7 Quantum complementarity} 

\medskip

Everywhere above we have been concerned with the fixed measurement
situation.

Certainly, for the different measurements one meets the different 
outcomes, which involve the different final states (e.g., $\Psi_i$
for the "object"), and the corresponding probabilities $W_i$. 
And those measurements which have identical outcomes, appear 
mutually equivalent in MTI. 

Here we meet the basis for "explanation" of the quantum complementarity
within MTI. The different outcomes define the {\it different,
mutually irreducible and incompatible distributions of the Time axes} .
Actually, due to {\bf (P1)} one can provide one, and only one (definite) 
Time axis
for an "object", and therefore {\it a definite measurement},
which refers to the actual Time-axes distribution. In other words :
{\bf (P1)} {\it implies mutual exclusiveness ("incompatibility") of the
different Time-axes distributions}, which is the MTI form of the
famous quantum complementarity.

\bigskip

{\bf 20.8 Macroscopic irreversibility}

\medskip

{\bf MTI implies macroscopic irreversibility !}

In order to prove this claim, we shall be concerned with an "object"
interacting with the macroscopic bodies. For the simplicity, we shall
be concerned only with the states of the "object", but everything
directly refers also to the states of the "apparatus" and "environment".

Let us consider a "hystory" of an "object"; this is a "chain" of
sequences of the free (Schrodinger) evolutions, and of the 
measurement-like interactions with the "microscopic" pieces of the
macroscopic bodies. And let us assume that the initial state of
the "object" is a "pure" state $\Psi$. Finally, let us assume that
the "object" interacts first with a macroscopic body $M_1$.

Then MTI leads to a (stochastic) transition of state of the "object" :
$$\Psi \rightarrow \chi_1. \eqno (29)$$

After this transition the "object" evolves according to the Schrodinger
law:
$$\chi_t = \hat U(t) \chi, \eqno (30)$$

\noindent
and let us assume that in an instant $t$, the "object" interacts
with another macroscopic body $M_2$. Then (another stochastic
transition):
$$\chi_t \rightarrow \kappa_2, \eqno (31)$$

\noindent
and so on.

Then the "hystory" of the "object" reads :
$$\Psi \buildrel T_1 \over \longrightarrow \chi_1
\buildrel T_M \over \longrightarrow \chi_t 
\buildrel  T'_2 \over \longrightarrow \kappa_2
\dots. \eqno (32)$$

{\bf And this "hystory" is not reversible.}

To see this, let us be concerned with the state $\kappa_2$ as
the initial state. Then, bearing in mind the macroscopic bodies
on the proper "places" in the "hystory", the inverse of (32)
reads :
$$\kappa_2 \buildrel T'_2 \over \longrightarrow \chi_t
\buildrel T_M \over \longrightarrow \chi_1 
\buildrel  T_1 \over \longrightarrow \Psi
\dots. \eqno (33)$$

Actually, MTI states that, e.g., when interacting with $M_2$,
the "object" can "meet" the axis $T'_2$, {\it but only with some 
probability} $W'_2$. If this would occur, then the transition from
$\chi_1$ to $\Psi$  - which would be due to $T_1$ - can also
be obtained {\it only} with some probability $W_1$. Therefore,
 an ensemble of the "objects" initially prepared in the state
 $\kappa_2$, would split, and {\it the hystory (33) appears admitable
 only with probabiluty }:
 $$W_1 \cdot W'_2. \eqno (34)$$
 
The expression (34) is easy to be generalized by :
$$\Pi_{i=1}^N W_i \to 0, \quad N \to \infty, \eqno (35)$$ 

\noindent
while 
$$\sum\limits_{i=1}^N W_i \neq 1, \eqno (36)$$

\noindent
due to the fact that, generally, 
the probabilities $W_i$ refer to the different,
mutually exclusive, probability distributions.

Needless to say, {\it the expressions} (35) and (36) {\it represent a formal
expression of the macroscopic irreversibility in MTI}.

\bigskip

{\bf 20.9 EPR correlations}

\medskip

Unfortunatelly, MTI does not have much to say about the "EPR
paradox".

Each EPR pair consists in the two microscopic (quantum) systems,
a system $S_1$, and $S_2$. Each of them, likewise the pair itself,
is an isolated quantum system.

According to MTI, since there is no "environment", {\it the pair 
can not meet the change of Time axis} - unless a measurement by
an "apparatus" is prepared. 

Therefore, the EPR pairs are {\it necessarily the objects
of the "macroscopic" Time}, while {\it all the conservation
laws being exactly fulfilled}. Now, MTI implies that the 
pairs evolve according to the Schrodinger equation, which
- as is well known - implies establising of the quantum correlations:
$$\Psi_1 \chi_2 \buildrel T_M \over \longrightarrow
\sum_i c_i \Psi_{1i} \chi_{2i}, \eqno (37)$$

\noindent
with obvious notation and $\langle \Psi_{1i}\vert \Psi_{1j}\rangle
= \delta_{ij}$,
$\langle \chi_{2i}\vert \chi_{2j}\rangle = \delta_{ij}$.

Now MTI implies that the pair, $S_1+S_2$, represents an object
of the evolution due to $T_M$, which {\it can not be told for the
subsystems}, $S_1$, and $S_2$. Actually, as it is distinguished by
d'Espagnat 1976, one can not ascribe a definite quantum state
to the subsystem's - rather, they are in mutual quantum 
correlations (the famous quantum nonseparability).

But {\it this is substantial matter in MTI}. Note : {\bf everywhere}
above, we defined the physical Time with regard to a {\it definite
state} of the actual system. {\it Without a definite state, MTI does
not refer a definite Time axis to the actual system}. Now, due
to {\bf (P1)} one {\it can say nothing about evolution of either
subsystem}, $S_1$, or $S_2$.

Finally, the pair as a whole is an object of definite Time axis,
and everything told above refers to each single pair, but not to
the subsystems ($S_1$, and/or $S_2$). And : the "locality" in this
context, which refer to interaction between $S_1$ and $S_2$ -
which do not have the definite Time axes - is not the {\it locality}
dealt with above, which refers to definite Time of the
actual system.

\bigskip

{\bf 20.10 Relation to the "objective reduction" of
Pernrose}

\medskip

In his recent book Penrose 1994 has pointed out a need for a new
paradigm, the so-called, "objective reduction (OR)", which would
prove to be a real physical process/effect allowing for what is
called the "state reduction process (collapse)" by von Neumann.

Here we just point out that the process of stochastic change of 
Time axis in MTI can be considered as a candidate for "OR" of
Penrose. For both "processes" have the same final efect, bearing
physical reality with respect to the elements of the corresponding 
Hilbert state space.

\bigskip

{\bf 21. COMPARISON OF MTI WITH SOME MEASUREMENT
THEORIES}

\bigskip

As a matter of fact, most of the theories/interpretations deal with 
the "object" - instead of the composite systems, O+A, and/or O+A+E.
But, as it was distingushed above, this approach proves to be somewhat
naive. Still, we shall compare the basics of MTI with some prominent
theories.

\bigskip

{\bf 21.1 MWI of Everett}

\medskip

It is probably obvious that MTI and MWI of Everett 1957 do 
not have very much in common. Still, there is a small danger for
misunderstanding the "channels" of MTI as the "branches" of MWI.
This is the main object of this Section.

As distinct from MTI, MWI deals with unique Time, and therefore with
the Schrodinger equation concerning (isolated)  O+A; i.e., there 
is no environment in this theory. Finally, the "branching" is considered
as {\it global, metaphysical process}, while the "channels" refer to
the {\it real, stochastic, local processes, which keep uniqueness of
the Universe}.

\bigskip

{\bf 21.2 Von Neumann's theory}

\medskip

Throughout the text we have strongly emphasized the distinctions
with this regard. As the most important distinction appears unnecessity
of the "state reduction process" (of the von Neumann's theory), which
actually becomes an unphysical process. And the measurement process
reduces onto the search for quantum effect, which would allow for the
local, stochastic change of Time.

{\bf 21.3 Some modern theories}

\medskip

There is a set of modern theories which bear mutual similarities,
at least as regards our goal - comparison with MTI. Particularly,
we mean the GRW theory of Ghirardi et al 1986, RPI theory of 
Mensky 1993, the method of stochastic Schrodinger equation of
Diosi 1989, likewise the recent interpretation (in the same context)
of Kist et al 1998.

All these theories are concerned with a search for a quantum law,
which would govern the evolution of the {\it "object" alone}; certainly,
everything expressed in $T_M$. And this is our impression that these
theories, altogether, narrow down the list of candidates for 
a general quantum law for the quantum "objects", the existence of which
has just been presumed - cf. Section 7 - but not considered in
detail.

Especially, the method of stochastic Schrodinger equation drags
attention. And especially the interpretation of Kist et al.
Actually, prima facie, this interpretation bears significant
similarities with MTI. Instead dealing with the details, we shall
just refer to the basic notions in this respect.

First, MTI deals with the composite systems, O+A and O+A+E. These
considerations justify the MTI statements concerning the "object" 
alone. Now, when expressed in terms of $T_M$, and with regard to
the "object" alone, MTI reduces onto the
notions raised in Section 7. And this really bears significant 
similarities with the interpretation of Kist et al. However, as
it follows from the Sections 12, this interpretation appears just as
a {\it model of MTI reduced onto} $T_M$. 

Actually, MTI does not insist on a particular type/form of the
operators $\hat U_i$ - cf. Sections 7. On the other side (cf.
Sections 7 and 12), the very existence of these operators - no
matter of which type - does not necessarily leads to MTI, but also to the
standard von Neumann's interpretation. Therefore, the 
interpretation of Kist et al, which deals only with the operators concerning
the "objects" alone, is {\it not equivalent with MTI}. Furthermore,
it can be considered only similar with a {\it particular
"reduction" of MTI onto the unique (universal) Time, and bearing in
mind the "object" alone}. 

\bigskip

{\bf 22. OUTLOOK}

\bigskip

MTI does not call for new hypotheses, but is rather a proposal for
new "re-reading" of the existing data concerning the quantum measurement
process.

MTI is {\it not} equivalent with any existing measurement theory
or interpretation.
On the other side, it is reducible - expressible - onto the terms
which refer to unique ("macroscopic") Time. To this end, 
the interpretation of
Kist et al 1998, appears as a {\it model of the "reduced MTI",
as regards the "object" alone}.

Still, MTI is consistent in both aspects, it is self-consistent, and
also consistent with - in so far as we can see - all the positive
statements of the quantum measurement theories.

Probably the main achievements of MTI appear to be {\it unnecessity
of the "state reductioin process"}, and {\it deducibility of the macroscopic
irreversibility}. This refers to the {\bf new physical picture}
concerning the "border teritory" between the "microscopic" and 
"macroscopic" parts of the World, given in Section 20. And, to
this end, it is very interesting to note : For rejecting the "reduction"
(as a necessary quantum process), it is necessary to have
the macroscopic (open quantum) systems as a part of the physical situation. 
On the other side, for deducing 
the macroscopic irreversibility, it is necessary to have the microscopic,
quantum systems involved. This points to the interplay between the "micro" and "macro"
in foundations of QM, and also points to the {\bf interaction} as a
fundamental issue in QM. [Eventually, the results of Dugi\' c 1996/7
can be of some interest in this concern.]

\vfill\eject

{\bf 23. CONCLUSION}

\bigskip

A particular, semi-epistemological definition of physical time
provides a {\bf new paradigm} in the quantum world. Particularly, in the
quantum measurement process, a single "object" becomes an object of 
stochastic change of Time (Time axis). Each Time should be 
considered objective
(real) for the actual "object", as the "macroscopic" Time is objective
in classical physics. The {\it possibility for changing Time refers to
 unique reference
frame}, and thus the introduced nonuniversality (nonuniqueness)
of Time does not have much in common with nonuniversality of Time
in the theory of relativity.

In the standard, ensemble-interpretation of QM, one meets the next
situation: in due course of the quantum measurement one meets
"indeterminism". As oposite to this, MTI keeps "determinism", but
with regard to the different Times. 

None existing theory/interpretation proves equivalent with MTI.
On the other side, MTI seems consistent with all the positive
statements of the existing measurement theories. This is the probably the 
basis
for obtaining the main achievements of MTI : {\it Rejecting the "state
reduction process" as a necessary (and actual) physical process},
and {\it Deducibility of the macroscopic irreversibility} - both trully 
ideals of each sound quantum-mecahnical theory. Within MTI, the 
"measurement problem" basically reduces onto the search for a 
physical mechanism (effect), which would allow for the 
local, stochastic change of
Time. Finally, we hope that this process can be recognized as the
"objective reduction" , as defined by Penrose 1994.

\bigskip

\bigskip

{\bf Appendix I}

\medskip

Within the Newton's theory of absolute time, the time "flows
equably without regard to anything external". Let us therefore
introduce the infinitesimal of this time, $dt_a$. Let us,
on the other side, introduce the infinitesimal of the physical
time measured by a clock - and, due to Newton, this time is not
identical with the absolute time - by $dt_c$. Generally speaking 
one may state :
$$dt_c = g(t_a) dt_a. \eqno(I.1)$$

Now, by {\it ideal clock} we mean any clock for which one may state
$g(t_a) = 1$.
I.e., there is (up to arbitrary additive constant) isomorphysm
between the two sets of instants (and therefore of the intervals) -
$t_a$ of the "absolute" time, and $t_{ci}$ of the "ideal clock".

From the {\bf purely operational physical point of view}, this admits
for considering {\it identity between the two Times} - the "absolute" one,
and the one measured by the "ideal clock". This identification is sometimes
expressed (cf. Mignard 1983)
by the phrase that "the modern 'atomic' clocks do both, 
'produce', {\it and} measure Time". 

This way one realizes that, from a purely {\it operational physical}
point of view, {\it physics can hardly ever say much more about Time,
than it is directly presented by the "ideal clock"}.

\bigskip

{\bf Appendix II}

\medskip

Both "primitives" of our approach - cf. subsection 2.2 - follow
from phenomenology - directly, or eventually by interpolation.

The dynamics $D$, Eq. (1), is actually an {\it elementary fact}
of physical observation, bearing no conditionality. This is certainly
the case with the {\it ordering of the elements} in $D$.

However, sometimes it is argued that, a priori, one can not make such
ordering without the prior ordering with respect to the time
instants. However, as it is well known - cf., e.g., Schuster 1961 -
in the, so-called, "relational theory of time", this is shown to be incorrect. 
Furthermore, as Penrose 1994 states, it is {\it just our psychology} that
inevitably calls for an a priori ordering in time. We shoud add : 
this a-priori-statement represents a "simultaneous" observation
{\it and} interpretation of the contents of observation.

Here, we advocate somewhat the oposite statement : we have the dynamics $D$,
and let us postpone the (above mentioned) interpretation. Actually, the
ordering in $D$ is {\it unavoidable physical fact}, which can be
(and usually is) recorded, memorized, If unconscious, the memory
contains a spatial ordering which can be
one-one re-written in terms of the points $A, B, \dots$ of the dynamics
$D$, without requireing an a priori temporal ordering.

\bigskip

{\bf Appendix III}

\medskip

Another "primitive" of our approach is the concept of "Causailty",
$C$.

This is also an element of the classical-physics phenomenology:
existence of unique and continuous "trajectory" in the state
space of the system.

Still, this has a deeper physical background: Actually, it is usually
(cf., e.g., Withrow 1979, p. 43) stated that there is a {\it close
connection between the (concept of) Time, and the} order {\it in the
physical world}.

The definition (3) {\it a priori} involves existence of the order.
Thus Time becomes a sort of "ordering principle" of the physical
world, rather than just being an (real) axis of the "time instants". 
Certainly, the epistemological concept of {\it "order"},
physically becomes the {\it "physical laws"}.

Still, the definition (3) should be considered somewhat {\it rudimentary}.
Actually, the definition (3) {\it establishes existence of Time},
without saying almost anything concerning the "nature of Time" 
(cf. Withrow 1979) -
which is still the subject of intense discussions and considerations;
cf. e.g., Zeh 1992. In other words : according to (3), the classical-physics
Time just exists, without involving the details concerning the diverse
discussions concerning the "nature of Time".

Let us finish by reminding that "Causality" is physically equivalent
with existence of physical laws, i.e., with the concepts of classical
determinism {\it and} causality (cf. Section 1 for definitions).

\bigskip

{\bf Appendix IV}

\medskip

The definition (3) essentially establishes equivalence of the concept
of physical Time, and physical dynamics, Eq. (2). This equivalence
can be justified as follows.

First, in the Newton's theory of time, the concept of physical
dynamics, $D_p$, naturally follows.

On the other side, the definition (3) gives a possibility for
stating physical definition of Time, Eq.(3). And this is logically
just inverse to the above statement. In this context, it is of
secondary importance if the instants in Eq. (3) are (non)unique.
However, {\it by definition}, we assume that if the instant
$t^{(M)}_A$ is fixed, than the instant $t^{(M)}_B$ should be considered
unique.

Therefore, the definition (3) establishes Time as a sort of "derivative"
of the physical dynamics.

Now, although following from the different backgrounds, the these 
two statements justify 
{\it physical equivalence of Time, and the physical dynamics}.

\bigskip

{\bf Appendix V}

\medskip

Rigorously speaking, existence of unique and continuous "trajectory"
concerning the i-th "Causality", $C_i$, is a matter of question,
which is here assumed valid, at least for the simplicity.

Still, we believe that this assumption is not necessary, but that
one may define some "generalizaed 'Causality' ", which would necessarily
lead to the definition of "generalized Time", thus opening the
question whether the microscopic Times, $T_i$, can be considered
{\it completely} physically equivalent with $T_M$.

\bigskip

{\bf Literature}

\bigskip

D' Espagnat, B., 1976 "Conceptual Foundations of Quantum Mechanics",
Reading , MA: Benjanim

Diosi, L. 1989, Phys. Rev. {\bf A40}, 1165

Dugi\' c, M., 1996, Physica Scripta {\bf 53}, 9

Dugi\' c, M., 1997, Physica Scripta {\bf 56}, 536

Dugi\' c, M. 1998, "Many Time Interpretation of the
Quantum Measurement Process", (Integral version),
unpublished

Everett, H., 1957, Rev. Mod. Phys. {\bf 29} 454

Ghirardi, G.-C. et al, 1986, Phys. Rev. {\bf D34}, 470

Kist, T.L.B. et al, 1998, Los Alamos E-prin Archive, quant-ph 
9805027 (unpublished)

Mignard, F., 1983 in "L'espace et le temps aujourd'hui/J.
Alegria...[et al]", Editions du Seuil, Sect. 9

Mensky, M. B., 1993, "Continuous Quantum Measurement and
Path Integrals", IOP, Bristol

Penrose, R., 1995 "Shadows of the Mind", Vintage

Schuster, M.M., 1961, Rev. Metaphys., {\bf 15}, 209

Von Neumann, J., 1955, "Mathematical Foundations of Quantum Mechanics",
Princeton

Withrow, G.J., 1979, "The Natural philosophy of Time",

Zeh, H.D., 1992, "The Physical Basis of the Direction of Time",
Springer-Verlag

Zurek, W.H., 1997, in "Foundations of Quantum Mechanics in the Light
of New Technology", Ed. S. Nakajima et al, World Scientific, Singapore;
p. 93

\end